\documentclass[mathleft
]{an}
\usepackage{graphicx}
\usepackage{times}
\begin{document}

\Pagespan{1}{}
\Yearpublication{2012}%
\Yearsubmission{2012}%
\Month{}%
\Volume{}%
\Issue{}%

\title{Radio observations of Planck clusters}

\author{Ruta Kale\inst{1,2}\fnmsep\thanks{Corresponding author:
  \email{rkale@ira.inaf.it}\newline}
}
\titlerunning{Radio observations of Planck clusters}
\authorrunning{Kale et al}
\institute{
INAF -- Istituto di Radioastronomia, via Gobetti 101, 40129 Bologna, Italy
\and Dipartimento di Astronomia, Universita di Bologna, via Ranzani 1, 40126 Bologna, Italy
}

\received{August 2012}
\accepted{September 2012}
\publonline{}


\abstract{Recently, a number of new galaxy clusters have been detected by the
ESA-Planck satellite,
the South Pole Telescope and the Atacama Cosmology Telescope using the
Sunyaev-Zel'dovich effect. Several of the newly detected clusters are massive,
merging systems with disturbed morphology in the X-ray surface brightness. Diffuse
 radio sources in clusters, called giant radio halos and relics, are direct
probes of
cosmic rays and magnetic fields in the intra-cluster medium. These radio sources
are found to occur mainly in massive merging clusters. Thus, the new
SZ-discovered clusters are good candidates to search for new radio halos and
relics. We have initiated radio observations of the clusters detected by Planck
with the Giant Metrewave Radio Telescope. These observations have already led to
the detection of a radio halo in PLCKG171.9-40.7, the first giant halo
discovered in one of the new Planck clusters.}

\maketitle

\section{Galaxy cluster surveys}
Galaxy clusters are the most massive (masses $\sim10^{14}-10^{15} M_{\odot}$) 
gravitationally bound objects in the present Universe, with dark matter 
dominating the gravity. The baryonic matter in clusters consists of 
hot ($10^{7}- 10^{8}$ K)  intra-cluster medium (ICM) and galaxies.
Large surveys in optical and X-ray bands have resulted in discoveries of 
thousands of clusters (eg. Abell et al 1980; Bohringer et al 2004). 
These surveys rely on the surface brightness of the emission from  
stars in the galaxies (optical band) and on the thermal X-ray emission from the 
ICM for the detections of clusters. These result in
 `flux-limited' catalogues of sources. Another signal 
that is used to detect clusters is the thermal Sunyaev Zel'dovich effect (SZ) - 
the spectral distortion of Cosmic Microwave Background (CMB) caused by inverse
 Compton scattering with the ICM (Sunyaev and Zel'dovich 1972). This signal is 
independent of redshift and the integrated thermal SZ effect is expected to 
trace cluster mass with low scatter (eg. Motl et al 2005). 
Thus SZ cluster surveys deliver mass limited catalogues to arbitrarily high 
redshift. Currently the South Pole Telescope (SPT, Carlstrom et al 2009), 
Atacama Cosmology Telescope (ACT, Fowler et al 2007) and {\it Planck Satellite} 
 (Planck collaboration 2011d) are surveying the sky in mm-waves to detect 
clusters using the SZ signal. 

There are a number of clusters discovered by these telescopes that 
have been confirmed by deep X-ray 
and optical observations. Notably, the {\em Planck} collaboration has 
published confirmation of 51 new galaxy clusters (Planck collaboration 2011a, 
2011b, 2011c, 2012a, 2012b). The validation of the cluster 
candidates was carried out by snapshot observations in X-rays with the 
{\em XMM Newton}. These newly discovered clusters were earlier missed by the
blind all 
sky X-ray surveys such as with the ROSAT due to their low surface brightness. 
This is evident from the fact that the new clusters have flat electron density 
profiles (Planck collaboration 2011a) as compared to the clusters that were
detected in surveys by the ROSAT. Thus apart from being massive, the newly 
discovered clusters using the SZ are likely to be merging systems with
 disturbed ICM.
\section{Radio observations of clusters}
Radio observations are the main probes of 
 non-thermal activities in galaxy clusters. They lead to discoveries of radio
galaxies related to 
active galactic nuclei (AGN), starburst galaxies and radio halos and 
relics in and around galaxy clusters. Knowledge of radio galaxies and 
starburst galaxies in galaxy clusters is important for learning about 
 the accretion and feedback which are fundamental to understand the state of 
the ICM.  
It is now well-known that, with the thermal ICM are mixed a population of 
relativistic particles (Lorentz factors $>1000$)
and magnetic fields ($\sim0.1-1\mu$G). Radio halos and relics are diffuse radio
sources 
associated with 
the relativistic electrons and magnetic fields  
in the ICM on $\sim$ Mpc scales. 

Radio halos are $\sim$ Mpc size 
radio sources that are typically located at cluster centers, cospatial with 
the X-ray emission. They are possibly related to the turbulence in the ICM
injected 
by cluster mergers (eg. Brunetti et al 2004) or to the amplified magnetic
fields 
due to merger (eg. Keshet 2010). The secondary electron models are also being 
considered, though have found less observational evidence (eg. Blasi \& 
Colafrancesco 1999, Pfrommer \& En\ss{}lin 2004, Keshet \& Loeb 2010).
Arc-like radio relics located toward 
peripheries of clusters are believed to be accelerated plasma at merger-shocks
(eg. Ensslin et al 1998). 

The known radio halos and arc-like 
relics in clusters are hosted by clusters that show signatures of mergers such 
as disturbed X-ray morphologies (Cassano et al 2010). 
Therefore, the newly detected clusters in SZ
surveys are promising targets for searching radio halos and relics.
Even among those that are not new but are among the strong detections 
of {\rm Planck} are massive clusters. Since some of these clusters do not have
deep high resolution X-ray observations currently, the signatures of merger 
are not established. There is also evidence that radio halos occur 
in massive, X-ray luminous and hot galaxy clusters (see Feretti et al 2012 
for a review and references therein). Thus, the Planck detected clusters 
and in general those detected using the SZ signal are likely to host 
radio halos and relics.

Majority of the observational studies of radio halos and relics have been  
carried out at 1.4 GHz band in which the VLA in its compact configurations 
(C and D) offered the best sensitivity. Radio halos and relics typically have 
synchrotron spectra with 
spectral indices $\alpha\sim -1.1$ to $-1.5$ ($S\propto\nu^{\alpha}$) -- 
steeper as compared to standard radio galaxies ($\alpha\sim -0.8$). The 
$\sim$ Mpc extents imply angular sizes of a few to several tens of arcminutes 
in nearby (z $\sim 0.4 - 0.02$) clusters. These characteristics make them 
suitable for low frequency ($<$ GHz) observations. In recent times, the 
Giant Metrewave 
Radio Telescope (GMRT) has proved to be an efficient instrument with 
its low frequency (150 -- 610 MHz) capabilities for studying radio halos and 
relics (eg. Venturi et al 2007, 
2008; Brunetti et al 2008; Giacintucci et al 2008; Macario et al 2010; 
Kale \& Dwarakanath 2010, 2012).
\section{GMRT observations of Planck clusters}
We have initiated a radio observations of galaxy clusters selected from the SPT
sample (Williamson et al. 2011) with the Australia Telescop Compact Array and
the Planck Early SZ (ESZ) catalogue (Planck collaboration 2011a) with the GMRT.
In this paper we discuss our work on the clusters selected from the ESZ.

The clusters detected in the ESZ catalogue were examined in the NRAO VLA 
Sky Survey (NVSS). The NVSS (Condon et al 1998) is a survey at 1.4 GHz with 
the VLA in D configuration of all the sky north of declination $-40^{\circ}$. 
It is sensitive to extended emission 
of the size of up to $15'$. Several of the presently known radio halos and relics 
have been discovered from the NVSS (eg. Giovannini, Tordi \& Feretti 1999
; Bagchi et al 2006). All the clusters that showed the presence of extended sources 
of size $\geq 500$ kpc were proposed for observation with the GMRT to confirm 
the nature of the extended emission (Table 1). As an example, we show here 
the NVSS map of the cluster PLCKG200.9-28.2 (Fig. 1). The elongated source at the 
southern edge of the X-ray emission does not have any obvious optical counterpart.
Its location at the edge of the cluster, along the same direction as the 
elongation in the X-ray distribution makes it a promising candidate for a radio relic. 

GMRT observations of 8 clusters were carried out in Cycle 21 (Oct. -- Nov. 2011)
in the dual frequency band (610, 235 MHz) and of PLCKG200.9-28.2 will be carried
out in Cycle 23 (Table 1). The clusters selected for GMRT observations
span a redshift range of
0.06 to 0.55. To detect extended structures of $\sim$1 Mpc extents at these
redshifts, a telescope with a capability of detecting angular scales $\sim 15' -
3'$ is required. In addition, sufficient resolution to separate point sources
from the extended emission is needed. The GMRT antennas are distributed in a
hybrid configuration that consists of a central compact array and an extended
Y-shaped array. At 610 and 235 MHz, it is possible to image angular scales up to
$\sim17'$ and $\sim44'$, respectively. The highest possible resolutions at 610
and 235 MHz are $\sim5''$ and $\sim13''$ (FWHM), respectively. With $\sim 5$ hr
observation of each field, sensitivities of $50 - 80$ $\mu$Jy beam$^{-1}$ at 610
MHz and $0.2 -0.4$ mJy beam$^{-1}$ are expected to be achieved.

The first outcome from these GMRT observations was the detection of 
a new giant radio halo in the cluster PLCKG171.9-40.7 (Plck171, hereafter) 
(Giacintucci et al 2012, submitted).

\begin{center}
\begin{table}
\caption{Candidates among the Planck clusters}
\begin{tabular}{lll}
\hline
Cluster           & z      &Planck \\ 
name              &        &$S/N$\\
\hline
\hline 
Abell 1795	      & 0.0622 &12.4 \\
Abell 0478            & 0.0882 &12.8 \\
Abell 2302	      & 0.1790 &6.5  \\
PLCKG171.94-40.65 & 0.2700 &10.6  \\
H1821+643	      & 0.2990 &6.9 \\
MACS J1731.6+2252     & 0.3890 &7.3 \\
RXCJ1206.2-0848       & 0.4414 &7.3 \\
MACS J1149.5+2223     & 0.5450 &7.1 \\
PLCKG200.9-28.2       & 0.22 &5.2 \\ 
\hline 
\end{tabular}\\
Note: The Planck signal to noise is reported from Planck collaboration 2011a,
2012a.
\end{table}
\end{center}

\begin{figure}
    \centering
    \includegraphics[height=8 cm]{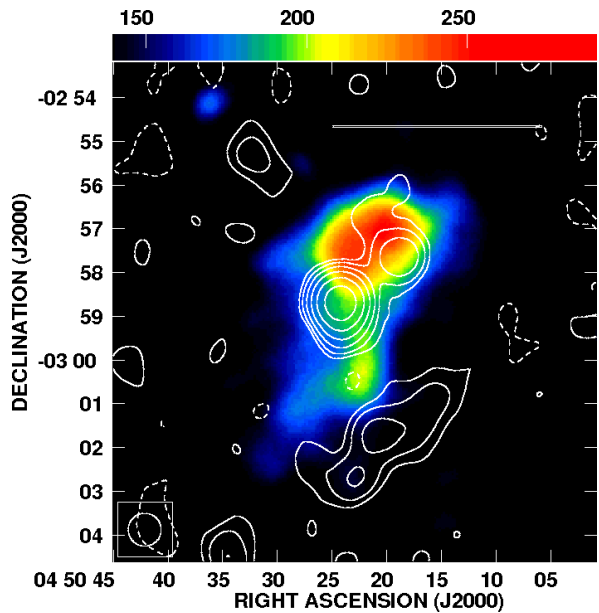}
    \caption{Plck200: Smoothed {\em XMM Newton} image (HEASARC) shown in colour
with 
contours of NVSS 
 (-1, 1, 2, 4, 8, 16, 32, 64 mJy beam$^{-1}$) overlaid. The horizontal segment denotes a 
distance of 1 Mpc 
at the redshift of Plck200. The elongated source located south of the X-ray 
emission is the suspected radio relic.} 
\end{figure}
\section{A radio halo in Plck171}
The cluster Plck171 is a newly detected cluster with the Planck and 
confirmed with the XMM Newton. It is 
located at RA 03h12m57.4s DEC +08d22m10s (J2000). It has a redshift of 
0.27 and X-ray luminosity (0.1 -- 2.4 keV) of $1.13\times10^{45}$ erg s$^{-1}$ 
(Planck collaboration 2011a). The analysis of {\em XMM Newton} data shows 
that the X-ray surface brightness is elongated in the 
northwest-southeast direction (Fig. 2, colour). The temperature of the 
cluster is estimated to be $\sim10$ keV. The X-ray morphology and temperature 
of the cluster suggest a merger in Plck171. 
\par The radio halo suspected from the NVSS images was confirmed with 
the GMRT 235 MHz observations (Fig. 2, contours) and a re-analysis of 
the NVSS data. The radio halo has a size of $\sim 1$ Mpc, 
typical of giant radio halos (eg. Feretti et al 2012).

\section{Discussion}
The discovery of new galaxy clusters with the SZ-effect shows that X-ray
telescopes are not sufficient to trace all the galaxy clusters. These newly
detected clusters affect the statistical studies that are based on samples of
galaxy clusters selected from X-ray flux limited catalogues. We illustrate this
point using the cluster Plck171. The cluster Plck171 is in the redshift and
X-ray luminosity range in which the GMRT Radio Halo Survey (GRHS: Venturi et al
2007, 2008) was carried out. 
The discovery of Plck 171 itself and further of the radio halo
can bias the statistics of the occurrence of radio halos obtained from
X-ray-selected cluster samples such as the GRHS. We will investigate this using 
the results obtained on the nine clusters listed here.

Correlations between the radio power of radio halo and cluster parameters such
as X-ray luminosity, temperature and
mass have been explored to find the clues to the origin of radio halos. 
In the X-ray luminosity ($L_x$) and radio power at 1.4 GHz ($P_{1.4GHz}$) plane,
 the clusters with radio halos and those without are well separated
 (Brunetti et al 2009). The clusters with radio halos also follow a scaling of 
radio power with X-ray luminosity (eg. Cassano et al 2008).  Recently, a
 correlation between 
the integrated SZ signal ($Y$) from the cluster and $P_{1.4GHz}$
 has been reported (Basu 2011). The scaling relations imply a connection
 between the thermal and non-thermal components in the clusters, if found
significant over a large range of parameters. The cluster Plck171 follows
both, the $L_x-P_{1.4GHz}$ and the $Y-P_{1.4GHz}$ correlations (Giacintucci et
al 2012, submitted).

In general, a large number of clusters need to be surveyed in order
to establish the statistical significance of the empirical correlations. 
 With the present observing capabilities, sensitive radio surveys of galaxy
clusters take long periods to reach completion. In future, surveys with
instruments such as the LOFAR (www.lofar.org) and LWA\\
(http://www.phys.unm.edu/~lwa/index.html) are expected to deliver sensitive
images of the sky at low frequencies (20 - 200 MHz). With the availability of
large statistical samples of galaxy clusters that are radio surveyed, it will be
possible to study the implications of the scaling relations and test the present
theoretical models.

\section{Summary}
Radio observations of Planck clusters, especially at low frequencies ($\leq$ 1.4 GHz), 
 are interesting to detect new radio halos and relics. Planck detected clusters that
 have possible diffuse radio emission in the NVSS 
images are being observed with the GMRT at 610 and 235 MHz. 
Eight clusters have already been observed and 
one more, with a suspected relic, will be observed. 
 Detection of a new radio halo in the cluster Plck171 is the first
 result from these observations. The analysis of data on other clusters is
ongoing. The results will be used to study the impact of these clusters on the
known statistical properties of radio halos and relics based on X-ray selected
cluster samples.

\begin{figure}
\includegraphics[width=8 cm]{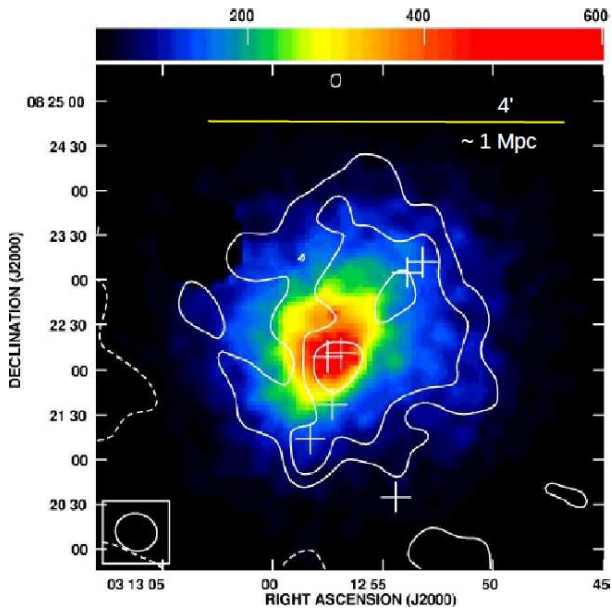}
\caption{GMRT 235 MHz image of the radio halo (contours) overlaid on 
XMM Newton X-ray image (Giacintucci et al. 2012, submitted). The $'+'$
symbols denote the locations of discrete 
radio sources from the 610 MHz map. These sources have been subtracted from 
the 235 MHz image.}
\label{plck171}
\end{figure}

\acknowledgements
We thank the anonymous referee for comments. RK thanks S. Giacintucci and M.
Markevitch for help during the preparation of this paper. We thank the staff
of the {\em
GMRT} for their help during the observations.
{\em GMRT} is run by the National Centre for Radio Astrophysics of the Tata
Institute of Fundamental Research. This research has made use of the NASA/IPAC 
Extragalactic Database (NED) which is operated by the Jet Propulsion 
Laboratory, California Institute of Technology, under contract with 
the National Aeronautics and Space Administration. This research has made use 
of data obtained from the High Energy Astrophysics Science Archive Research 
Center (HEASARC), provided by NASA's Goddard Space Flight Center.

\newpage

\end{document}